\documentclass[12pt]{article}
\usepackage[dvips]{color}
\usepackage{epsfig}
\usepackage{amsmath}
\usepackage{graphicx}

\begin{document}
\title{\bf Energy Loss of a Heavy Particle near $3D$ Charged Rotating Hairy Black Hole}
\author{{Jalil Naji\thanks{Email:
Naji{\_}jalil@yahoo.com}}\\
{\small {\em Physics Department, Ilam University, Ilam, Iran}}\\
{\small {\em P.O.Box 69315-516, Ilam, Iran}}} \maketitle
\begin{abstract}
In this paper we consider charged rotating black hole in
3 dimensions with an scalar charge and discuss about energy loss of heavy particle moving near the black hole horizon. We also study quasi-normal modes and find dispersion relations. We find that the effect of scalar charge and electric charge is increasing energy loss.
\end{abstract}

\section{Introduction}
The lower dimensional theories
may be used as toy models to study some fundamental ideas which yield to better understanding of
higher dimensional theories, because they are easier to study [1].
Moreover, these are useful for application of AdS/CFT correspondence [2-5]. This paper is indeed an application of AdS/CFT correspondence to probe moving charged particle near the three dimensional black holes which recently introduced by the Refs. [6] and [7] where charged black hole with a scalar hair in (2+1) dimensions, and rotating hairy black hole in (2+1)
dimensions constructed respectively.  Here we are interested to the case of rotating black hole with a scalar hair in (2+1) dimensions. Recently, a charged rotating hairy black hole in 3 dimensions corresponding to infinitesimal black hole parameters constructed [8] which will be used in this paper. Also thermodynamics of such systems recently studied by the Refs. [9] and [10].\\
In this work we would like to study motion of a heavy charged particle near the black hole horizon and calculate energy loss. The energy loss of moving heavy charged particle through
a thermal medium known as the drag force. One can consider a moving
heavy particle (such as charm and bottom quarks) near the black hole horizon with the momentum $P$, mass $m$ and constant velocity $v$,
which is influenced by an external force $F$. So, one can write the
equation of motion as $\dot{P}=F-\zeta P$, where in the
non-relativistic motion $P=mv$, and in the relativistic motion
$P=mv/\sqrt{1-v^{2}}$, also $\zeta$ is called friction coefficient.
In order to obtain drag force, one can consider two special cases.
The first case is the constant momentum which yields to obtain $F=(\zeta m)v$ for the non-relativistic case. In this case
the drag force coefficient $(\zeta m)$ will be obtained. In the
second case, external force is zero, so one can find
$P(t)=P(0)exp(-\zeta t)$. In another word, by measuring the ratio
$\dot{P}/P$ or $\dot{v}/v$ one can determine friction coefficient
$\zeta$ without any dependence on mass $m$. These methods lead us to
obtain the drag force for a moving heavy particle. The moving heavy particle in context of QCD has dual picture in
the string theory in which an open string attached to the D-brane and
stretched to the horizon of the black hole.\\
Similar studies already performed in several backgrounds [11-22]. Now, we are going to consider the same problem in a charged rotating hairy 3D background. Our motivation for this study is $AdS_{3}/CFT_{2}$ correspondence [23-25].\\
This paper is organized as the following. In the next section we review charged rotating hairy black hole in (2+1) dimensions. In section 3 we obtain equation of motion and in section 4 we try to obtain solution and discuss about energy loss. In section 5 we give linear analysis and discuss about quasi-normal modes and dispersion relations. Finally in section 6 we summarized our results and give conclusion.

\section{Charged rotating hairy black hole in (2+1) dimensions}
The (2+1)-dimensional gravity with a non-minimally
coupled scalar field is
described by the following action,
\begin{equation}\label{s1}
S=\frac{1}{2}\int{d^{3}
x\sqrt{-g}[R-g^{\mu\nu}\nabla_{\mu}\phi\nabla_{\nu}\phi-\xi
R\phi^2-2V(\phi)-\frac{1}{4}F_{\mu\nu}F^{\mu\nu}]},
\end{equation}
where $\xi$ is a coupling constant between gravity and the scalar
field which will be fixed as $\xi=1/8$, and $V(\phi)$ is self coupling potential. The metric background of this
is given by the Ref. [7],
\begin{eqnarray}\label{s2}
ds^{2}=-f(r)dt^{2}+\frac{1}{f(r)}dr^{2}+r^{2}(d\psi+\omega(r)dt)^{2},
\end{eqnarray}
where [8],
\begin{equation}\label{s3}
f(r)=3\beta-\frac{Q^2}{4}+(2\beta-\frac{Q^2}{9})\frac{B}{r}-Q^2(\frac{1}{2}+\frac{B}{3r})\ln(r)+\frac{(3r+2B)^{2}a^{2}}{r^{4}}+\frac{r^2}{l^2}+\mathcal{O}(a^{2}Q^{2}).
\end{equation}
where $Q$ is infinitesimal electric charge, $a$ is infinitesimal rotational parameter and $l$ is related
to the cosmological constant as $\Lambda=-\frac{1}{l^{2}}$. Also $\beta$
is integration constants depends on the black hole charge and mass as the following,
\begin{equation}\label{s4}
\beta=\frac{1}{3}(\frac{Q^2}{4}-M),
\end{equation}
and the scalar charge $B$ related to the scalar field as the following,
\begin{equation}\label{s5}
\phi(r)=\pm\sqrt{\frac{8B}{r+B}}.
\end{equation}
Rotational frequency obtained as the following,
\begin{equation}\label{s6}
\omega(r)=-\frac{(3r+2B)a}{r^{3}},
\end{equation}
and,
\begin{equation}\label{s7}
V(\phi)=\frac{2}{l^{2}}+\frac{1}{512}\left[\frac{1}{l^{2}}+\frac{\beta}{B^{2}}+\frac{Q^{2}}{9B^{2}}\left(1-\frac{3}{2}\ln(\frac{8B}{\phi^{2}})\right)\right]\phi^{6}
+\mathcal{O}(Q^{2}a^{2}\phi^{8}).
\end{equation}
Also one can obtain the following Ricci scalar,
\begin{equation}\label{s8}
R=-\frac{36r^{6}-3l^{2}Q^{2}r^{4}+2Bl^{2}Q^{2}r^{3}+216Bl^{2}a^{2}r+180l^{2}a^{2}B^{2}}{6l^{2}r^{6}},
\end{equation}
which is singular at $r = 0$. Finally, in the Ref. [8] it is found that,
\begin{equation}\label{s9}
r_{h}^{2}=\frac{B}{3Q^{2}}(\frac{7}{6}Q^{2}-2M)\left(-1+\sqrt{1+\frac{216a^{2}Q^{2}}{B(\frac{7}{6}Q^{2}-2M)^{2}}}\right),
\end{equation}
where $r_{h}$ is the black hole horizon radius.\\
Finally black hole temperature and entropy are obtained by the following relations,
\begin{eqnarray}\label{s10}
T&=&\frac{f^{\prime}(r_{h})}{4\pi},\nonumber\\
s&=&4\pi r_{h}.
\end{eqnarray}
\section{The equations of motion}
The moving heavy particle near the black hole may be described by the following Nambu-Goto action,
\begin{equation}\label{s11}
S=-\frac{1}{2\pi\alpha^{\prime}}\int{d\tau d\sigma \sqrt{-G}},
\end{equation}
where $T_{0}=\frac{1}{2\pi\alpha^{\prime}}$ is the string tension. The coordinates $\tau$ and $\sigma$ are corresponding to the string world-sheet. Also $G_{ab}$
is the induced metric on the string world-sheet with determinant $G$ obtained as the following,
\begin{equation}\label{s12}
G=-1-r^{2}f(r)(x^{\prime})^{2}+\frac{r^{2}}{f(r)}(\dot{x})^{2},
\end{equation}
where we used static gauge in which
$\tau = t$ , $\sigma = r$, and the string only extends in one direction $x(r, t)$. Then, the equation of motion obtained as the following,
\begin{equation}\label{s13}
\partial_{r}\left(\frac{r^{2}f(r)x^{\prime}}{\sqrt{-G}}\right)-\frac{r^{2}}{f(r)}\partial_{t}\left(\frac{\dot{x}}{\sqrt{-G}}\right)=0.
\end{equation}
We should obtain canonical momentum densities associated to the string as the follows,
\begin{eqnarray}\label{s14}
\pi_{\psi}^{0}&=&\frac{1}{2\pi\alpha^{\prime}\sqrt{-G}}\frac{r^{2}}{f(r)}\dot{x},\nonumber\\
\pi_{r}^{0}&=&-\frac{1}{2\pi\alpha^{\prime}\sqrt{-G}}\frac{r^{2}}{f(r)}\dot{x}x^{\prime},\nonumber\\
\pi_{t}^{0}&=&-\frac{1}{2\pi\alpha^{\prime}\sqrt{-G}}(1+r^{2}f(r)(x^{\prime})^{2}),\nonumber\\
\pi_{\psi}^{1}&=&\frac{1}{2\pi\alpha^{\prime}\sqrt{-G}}r^{2}f(r)x^{\prime},\nonumber\\
\pi_{r}^{1}&=&-\frac{1}{2\pi\alpha^{\prime}\sqrt{-G}}(1-\frac{r^{2}}{f(r)}\dot{x}^{2}),\nonumber\\
\pi_{t}^{1}&=&\frac{1}{2\pi\alpha^{\prime}\sqrt{-G}}r^{2}f(r)\dot{x}x^{\prime}.
\end{eqnarray}
The simplest solution of the equation of motion is static string described by $x=const.$ with total energy of the form,
\begin{equation}\label{s15}
E=-\int_{r_{h}}^{r_{m}}{\pi_{t}^{0}dr}=\frac{1}{2\pi\alpha^{\prime}}(r_{h}-r_{m})=M_{rest},
\end{equation}
where $r_{m}$ is arbitrary location of D-brane. As we expected, the energy of static particle interpreted as rest mass.
\section{Time dependent solution}
In the general case, we can assume that particle moves with constant speed $\dot{x}=v$, in that case the equation of motion (13) reduces to,
\begin{equation}\label{s16}
\partial_{r}\left(\frac{r^{2}f(r)x^{\prime}}{\sqrt{-G}}\right)=0,
\end{equation}
where,
\begin{equation}\label{s17}
G=-1-r^{2}f(r)(x^{\prime})^{2}+\frac{r^{2}}{f(r)}v^{2}.
\end{equation}
The equation (16) gives the following expression,
\begin{equation}\label{s18}
(x^{\prime})^{2}=\frac{C^{2}(r^{2}v^{2}-f(r))}{r^{2}f(r)^{2}(C^{2}-r^{2}f(r))},
\end{equation}
where $C$ is an integration constant which will be determined by using reality condition of $\sqrt{-G}$. Therefore we yield to the following canonical momentum densities,
\begin{eqnarray}\label{s19}
\pi_{\psi}^{1}&=&-\frac{1}{2\pi\alpha^{\prime}}C,\nonumber\\
\pi_{t}^{1}&=&\frac{1}{2\pi\alpha^{\prime}}Cv.
\end{eqnarray}
These give us loosing energy and momentum through an
endpoint of string,
\begin{eqnarray}\label{s20}
\frac{dP}{dt}=\pi_{\psi}^{1}|_{r=r_{h}}&=&-\frac{1}{2\pi\alpha^{\prime}}C,\nonumber\\
\frac{dE}{dt}=\pi_{t}^{1}|_{r=r_{h}}&=&\frac{1}{2\pi\alpha^{\prime}}Cv.
\end{eqnarray}
As we mentioned before, reality condition of $\sqrt{-G}$ gives us constant $C$. The expression $\sqrt{-G}$ is real for $r=r_{c}>r_{h}$. In the case of small $v$ one can obtain,
\begin{equation}\label{s21}
r_{c}=r_{h}+\frac{r^{2}v^{2}}{f(r)^{\prime}}|_{r=r_{h}}+\mathcal{O}(v^{4}),
\end{equation}
which yields to,
\begin{equation}\label{s22}
C=vr_{h}^{2}+\mathcal{O}(v^{3}).
\end{equation}
Therefore we can write drag force as the following,
\begin{equation}\label{s23}
\frac{dP}{dt}=-\frac{vr_{h}^{2}}{2\pi\alpha^{\prime}}+\mathcal{O}(v^{3}).
\end{equation}
In the Fig. 1 we can see behavior of drag force with the black hole parameters. We draw drag force in terms of velocity and as expected, value of drag force increased by $v$. Fig. 1 (a) and (b) show that the black hole electric charge as well as scalar charge increase value of drag force. We find a lower limit for the black hole charge which is for example $Q\geq1.4$ corresponding to $M=a=B=1$. In this case we find that slow rotational motion has many infinitesimal effect on drag force which may be negligible.
\begin{figure}[th]
\begin{center}
\includegraphics[scale=.3]{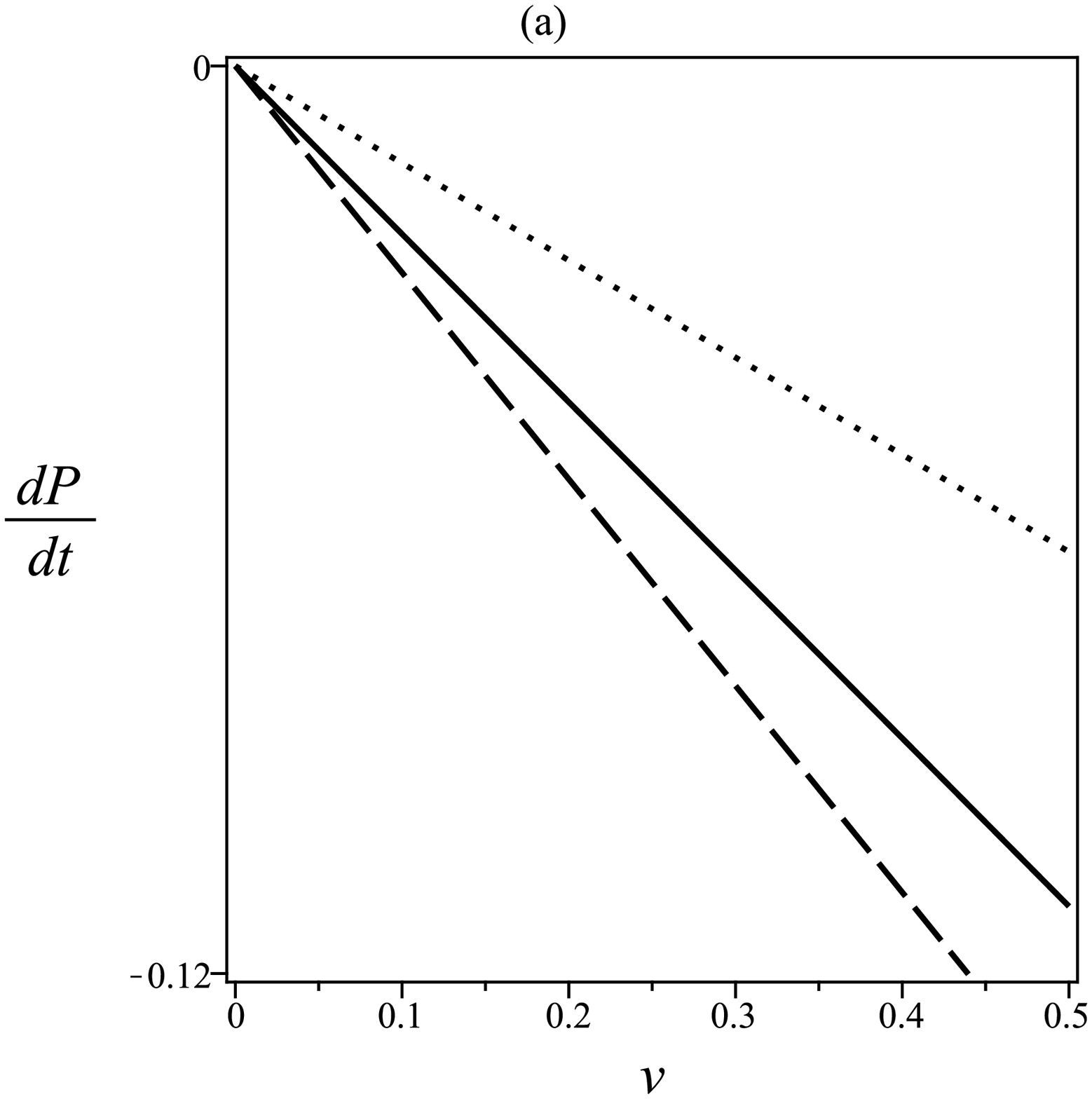}\includegraphics[scale=.3]{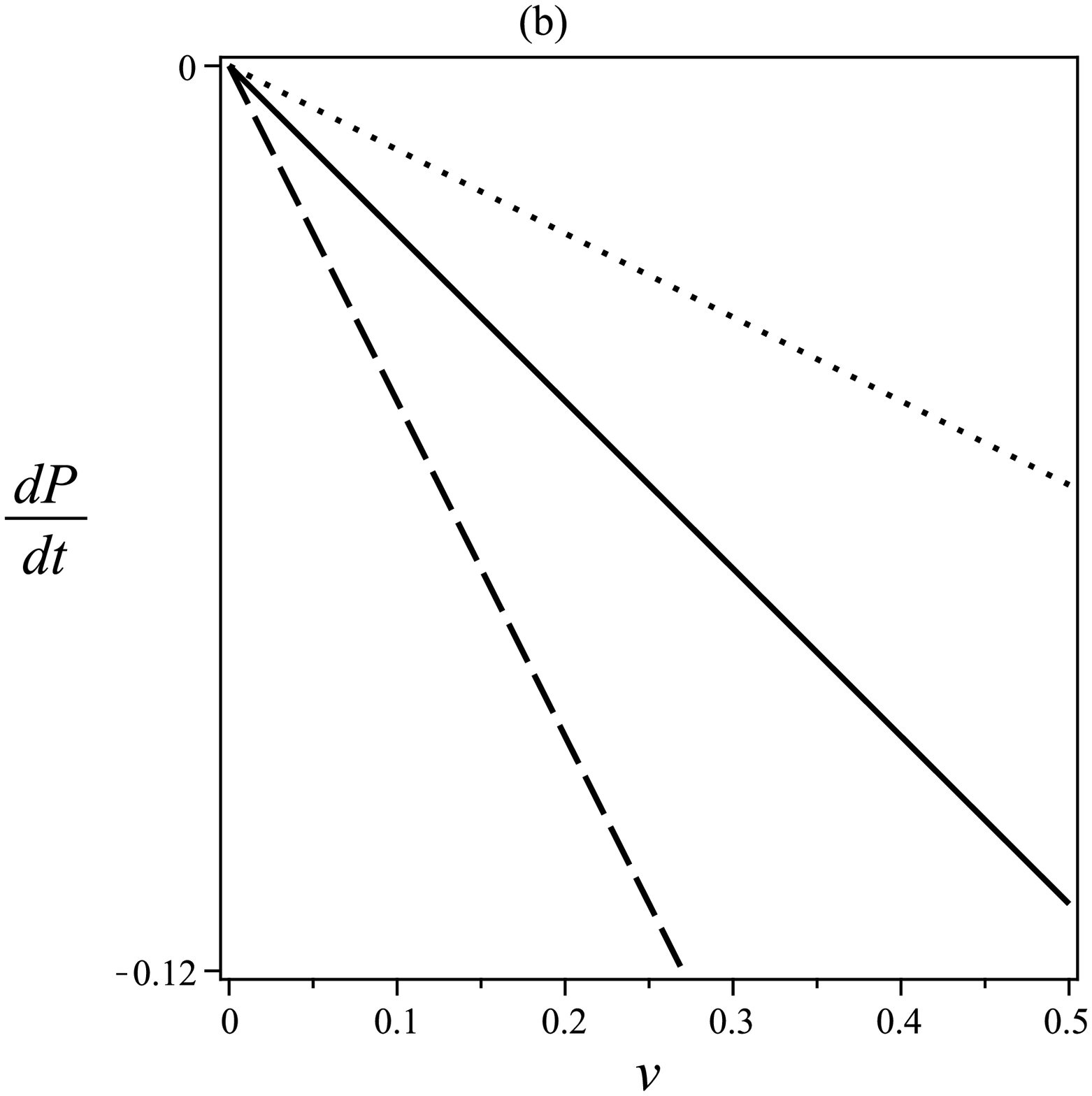}
\caption{Drag force in terms of $v$ for $M=1$. (a) $B=a=1$, $Q=1.6$ (dotted line), $Q=2$ (solid line), $Q=2.4$ (dashed line). (b) $a=1$, $Q=2$, $B=0.5$ (dotted line), $B=1$ (solid line), $B=2$ (dashed line).}
\end{center}
\end{figure}

\section{Linear analysis}
Because of drag force, motion of string yields to small perturbation after late time. In that case speed of particle is infinitesimal and one can write $G\approx-1$. Also we assume that $x=e^{-\mu t}$, where $\mu$ is the friction coefficient. Therefore one can rewrite the equation of motion as the following,
\begin{equation}\label{s24}
\frac{f(r)}{r^{2}}\partial_{r}(r^{2}f(r)x^{\prime})=\mu^{2}x.
\end{equation}
We assume out-going boundary conditions near the black hole horizon and use the following approximation,
\begin{equation}\label{s25}
(4\pi T)^{2}(r-r_{h})\partial_{r}(r-r_{h})x^{\prime}=\mu^{2}x.
\end{equation}
which suggest the following solutions,
\begin{equation}\label{s26}
x=c(r-r_{h})^{-\frac{\mu}{4\pi T}},
\end{equation}
where $T$ is the black hole temperature. In the case of infinitesimal $\mu$ we can use the following expansion,
\begin{equation}\label{s27}
x=x_{0}+\mu^{2}x_{1}+...
\end{equation}
Inserting this equation in the (25) gives $x_{0}=const.$, and,
\begin{equation}\label{s28}
x_{1}^{\prime}=\frac{A}{r^{2}f(r)}\int_{r_{h}}^{r_{m}}\frac{r^{2}}{f(r)}dr,
\end{equation}
where $A$ is a constant. Assuming near horizon limit enables us to obtain the following solution,
\begin{equation}\label{s29}
x_{1}\approx\frac{A}{4\pi T r_{h}^{2}(r-r_{h})}\left(-r_{m}+\frac{r_{h}^{2}}{4\pi T}\ln{(r-r_{h})}\right).
\end{equation}
Comparing (26) and (28) gives the following quasi-normal mode condition,
\begin{equation}\label{s30}
\mu=\frac{r_{h}^{2}}{r_{m}}.
\end{equation}
It is interesting to note that these results recover drag force (23) for infinitesimal speed. In the Fig. 2 we can see behavior of $\mu$ with the black hole parameters. We find that black hole charges increase value of friction coefficient.

\begin{figure}[th]
\begin{center}
\includegraphics[scale=.3]{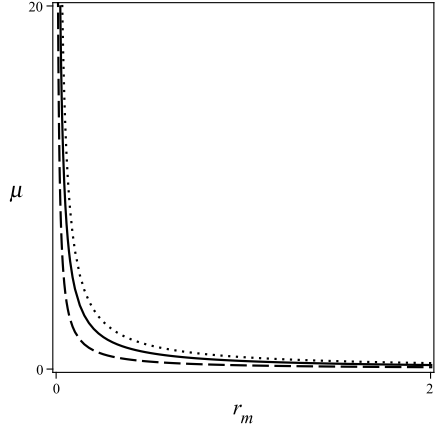}
\caption{$\mu$ in terms of $r_{m}$. $B=Q=a=1.6$ (dashed line), $B=Q=a=2$ (solid line), $B=Q=a=2.4$ (dotted line).}
\end{center}
\end{figure}

\subsection{Low mass limit}
Low mass limit means that $r_{m}\rightarrow r_{h}$, and we use the following assumptions,
\begin{equation}\label{s31}
f(r)\approx4\pi T(r-r_{h}),
\end{equation}
and,
\begin{equation}\label{s32}
r^{2}=r_{h}^{2}+2r_{h}(r-r_{h})+... ,
\end{equation}
so, by using the relation (24) we can write,
\begin{equation}\label{s33}
x(r)=(r-r_{h})^{-\frac{\mu}{4\pi T}}(1+(r-r_{h})A+...).
\end{equation}
We can obtain constant $A$ as the following,
\begin{equation}\label{s34}
A=\frac{\mu}{2\pi T r_{h}-\mu r_{h}}.
\end{equation}
It tells that $\mu=2\pi T$ yields to divergency, therefore we called this as critical behavior of the friction coefficient and obtain Fig. 3.

\begin{figure}[th]
\begin{center}
\includegraphics[scale=.3]{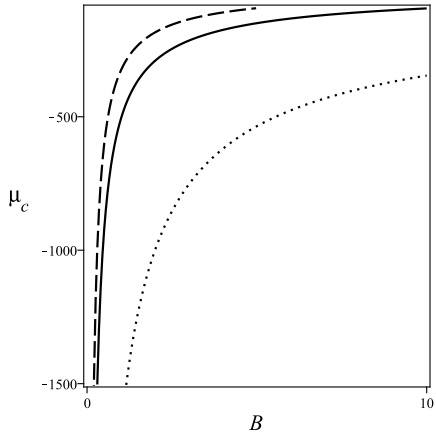}
\caption{$\mu_{c}$ in terms of $B$ for $M=1$, $l=1$ and $a=2$. $Q=1.6$ (dotted line), $Q=2$ (solid line), $Q=2.4$ (dashed line).}
\end{center}
\end{figure}

\subsection{Dispersion relations}
Here, we would like to obtain relation between total energy $E$ and momentum $P$ at the slow velocity limit. In that case we can obtain,
\begin{equation}\label{s35}
\pi_{\psi}^{0}=-\frac{\mu}{2\pi\alpha^{\prime}}\frac{r^{2}}{f(r)}=-\frac{1}{2\pi\alpha^{\prime}\mu}\partial_{r}(r^{2}f(r)x^{\prime}),
\end{equation}
which gives the total momentum,
\begin{equation}\label{s36}
P=\int\pi_{\psi}^{0}dr=\frac{1}{2\pi\alpha^{\prime}\mu}\left(r_{min}^{2}f(r_{min})x^{\prime}(r_{min})\right),
\end{equation}
where we use $r_{min}>r_{h}$ as IR cutoff to avoid divergency. In the similar way we can compute another momentum density,
\begin{equation}\label{s37}
\pi_{t}^{0}=-\frac{\mu}{2\pi\alpha^{\prime}}\left[1+\frac{1}{2}r^{2}f(r)(x^{\prime})^{2}+\frac{1}{2}\frac{r^{2}}{f(r)}v^{2}\right],
\end{equation}
to evaluate total energy as the following,
\begin{equation}\label{s38}
E=-\int\pi_{t}^{0}dr=\frac{1}{2\pi\alpha^{\prime}}\left(r_{m}-r_{min}-r_{min}^{2}f(r_{min})x^{\prime}(r_{min})x(r_{min})\right),
\end{equation}
where we used equation of motion and boundary condition $x^{\prime}(r_{m})=0$. Assuming the following near horizon solution,
\begin{equation}\label{s39}
x\sim (r-r_{h})^{-\frac{\mu}{4\pi T}}
\end{equation}
and combining equations (36) and (38) give the following relation,
\begin{equation}\label{s40}
E=M_{rest}+\frac{P^{2}}{2M_{kin}},
\end{equation}
where $M_{rest}$ given by the equation (15) with replacement $r_{h}\rightarrow r_{min}$ and,
\begin{equation}\label{s41}
M_{kin}\equiv\frac{r_{h}^{2}}{2\pi \alpha^{\prime}\mu}.
\end{equation}
It is usual, non-relativistic
dispersion relation for a point particle in which the rest mass is different
from the kinetic mass. In the Fig. 4 we draw re-scaled $\eta\equiv 2\pi \alpha^{\prime}\mu$ in terms of kinetic mass and show that black hole charges increase $\eta$ as expected.
\begin{figure}[th]
\begin{center}
\includegraphics[scale=.3]{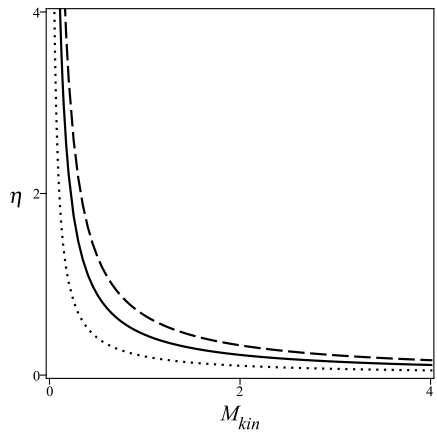}
\caption{$\eta$ in terms of $M_{kin}$. $Q=B=1.6$ (dotted line), $Q=B=2$ (solid line), $Q=B=2.4$ (dashed line).}
\end{center}
\end{figure}

\section{Conclusions}
In this work we considered recently constructed charged rotating black hole in 3 dimensions with an scalar
charge and calculated energy loss of heavy particle moving near the black hole horizon. First of all important properties of background reviewed and then appropriate equations obtained. We use motivation of AdS/CFT correspondence and use string theory method to study motion of particle. This is indeed in the context of $AdS_{3}/CFT_{2}$ where drag force on moving heavy particle calculated. We found that the black hole charges, both electric and scalar, increase value of drag force but infinitesimal value of rotation parameter has no any important effect and may be negligible. We also discussed about quasi-normal modes and obtained friction coefficient and found that black hole
charges increase value of friction coefficient which is coincide with increasing drag force. Finally we found dispersion relation which relates total energy and momentum of particle.


\begin{thebibliography}{11}
\bibitem{P1}
M. Hortacsu, H.T. Ozcelik, B. Yapiskan, "Properties of Solutions in
2+1 Dimensions", Gen. Rel. Grav. 35 (2003) 1209
\bibitem{P2}
M. Henneaux, C. Martinez, R. Troncoso, and J. Zanelli, "Black holes
and asymptotics of 2+1 gravity coupled to a scalar field", Phys.Rev.
D65 (2002) 104007
\bibitem{P3}
M. Hasanpour, F. Loran, and H. Razaghian, "Gravity/CFT
correspondence for three dimensional Einstein gravity with a
conformal scalar field",  Nucl. Phys. B867 (2013) 483
\bibitem{P4}
D.F. Zeng, "An Exact Hairy Black Hole Solution for AdS/CFT
Superconductors", [arXiv:0903.2620 [hep-th]].
\bibitem{P5}
B. Chen, Z. Xue, and J. ju Zhang, "Note on Thermodynamic Method of
Black Hole/CFT Correspondence", JHEP 1303 (2013) 102
\bibitem{P6}
W. Xu, L. Zhao, "Charged black hole with a scalar hair in (2+1)
dimensions", Phys. Rev. D87 (2013) 124008
\bibitem{P7}
L. Zhao, W. Xu, B. Zhu, "Novel rotating hairy black hole in
(2+1)-dimensions", [arXiv:1305.6001 [gr-qc]]
\bibitem{P8}
J. Sadeghi, B. Pourhassan, H. Farahani, "Rotating charged hairy black hole in (2+1) dimensions and particle acceleration", [arXiv:1310.7142 [hep-th]]
\bibitem{P9}
A. Belhaj, M. Chabab, H. EL Moumni, M. B. Sedra, "Critical Behaviors
of 3D Black Holes with a Scalar Hair", [arXiv:1306.2518 [hep-th]]
\bibitem{P10}
J. Sadeghi, H. Farahani, "Thermodynamics of a charged hairy black
hole in (2+1) dimensions", [arXiv:1308.1054 [hep-th]]
\bibitem{P11}
Carlos Hoyos-Badajoz, "Drag and jet quenching of heavy quarks in a
strongly coupled N=2* plasma", JHEP 0909(2009) 068, [arXiv:0907.5036
[hep-th]].
\bibitem{P12}
J. Sadeghi and B. Pourhassan, " Drag force of moving quark at the
${\mathcal{N}} =2$ supergravity", JHEP 0812 (2008) 026,
[arXiv:0809.2668 [hep-th]].
\bibitem{P13}
C. P. Herzog, A. Karch, P. Kovtun, C. Kozcaz, and L. G. Yaffe,
"Energy loss of a heavy quark moving through ${\mathcal{N}} =4$
supersymmetric Yang-Mills plasma" JHEP 0607 (2006) 013, [arXiv:
hep-th/0605158].
\bibitem{P14}
J. Sadeghi, B. Pourhassan and S. Heshmatian, "Application of AdS/CFT in quark-gluon plasma", Advances in High Energy Physics 2013 (2013) 759804
\bibitem{P15}
C.P. Herzog, "Energy loss of heavy quarks from asymptotically AdS
geometries", JHEP 0609 (2006) 032, [arXiv: hep-th/0605191].
\bibitem{P16}
S.S. Gubser, "Drag force in AdS/CFT", Phys. Rev. D74 (2006) 126005.
\bibitem{P17}
E. Nakano, S. Teraguchi and W.Y. Wen, "Drag Force, Jet Quenching,
and AdS/QCD", Phys. Rev. D 75 (2007) 085016.
\bibitem{P18}
E. Caceres and A. Guijosa, "Drag force in charged ${\mathcal{N}}=4$
SYM plasma". JHEP 0611 (2006) 077.
\bibitem{P19}
J.F. Vazquez-Poritz, "Drag force at finite 't Hooft coupling from
AdS/CFT", [arXiv: hep-th/0803.2890].
\bibitem{P20}
A.N. Atmaja and K. Schalm, "Anisotropic Drag Force from 4D Kerr-AdS
Black Holes", [arXiv:1012.3800 [hep-th]].
\bibitem{P21}
B. Pourhassan  and J. Sadeghi "STU/QCD correspondence", [arXiv:1205.4254 [hep-th]] Can J Phys
\bibitem{P22}
E. Caceres and A. Guijosa, "On drag forces and jet quenching in
strongly coupled plasmas", JHEP 0612 (2006) 068.
\bibitem{P23}
P. Kraus, "Lectures on Black Holes and the AdS3/CFT2 Correspondence", Supersymmetric Mechanics - Vol. 3
Lecture Notes in Physics 755 (2008) 1
\bibitem{P24}
R. Borsato, O.O. Sax, A. Sfondrini, "All-loop Bethe ansatz equations for AdS3/CFT2", JHEP 1304 (2013) 116
\bibitem{P25}
D. Momeni, M. Raza, M.R. Setare, R. Myrzakulov, "Analytical Holographic Superconductor with Backreaction Using AdS3/CFT2", International Journal of Theoretical Physics 52 (2013)  2773

\end{thebibliography}
\end{document}